\documentclass[epjCONF]{svjour}
\usepackage{graphics}
\session-title{A Universe of Dwarf Galaxies}
\begin{document}
\title{3D spectroscopy of dwarf elliptical galaxies}
\author{Olga K. Sil'chenko\inst{1}\fnmsep\thanks{\email{olga@sai.msu.su}}}
\institute{Sternberg Astronomical Institute, Moscow, Russia}
\abstract{
I present some results of 3D spectroscopy for a small sample of dwarf
elliptical galaxies, mostly members of small groups. The galaxies
under consideration have a typical absolute magnitude of $-18$ ($B$), and at the
Kormendy's relation they settle within a transition zone between the
main cloud of giant ellipticals and the sequence of diffuse ellipticals.
By measuring Lick indices and investigating radial profiles of the SSP-equivalent
ages and metallicities of the stellar populations in their central parts, I have 
found evolutionary distinct cores in all of them. Typically, the ages of these
cores are 2-4 Gyr, and the metallicities are higher than the solar one.
Outside the cores, the stellar populations are always old, $T>12$ Gyr, and
the metallicities are subsolar. This finding implies that the well-known
correlation between the stellar age and the total mass (luminosity) of
field ellipticals \cite{trager00,caldwell,howell}
may be in fact a direct consequence of a larger contribution of nuclear
starbursts into the integrated stellar population in dwarfs with respect
to giants, and does not relate to `downsizing'. 
} 
\maketitle
\section{Introduction}
\label{intro}
The results of panoramic spectroscopy and spatially resolved
parameters of the stellar populations (ages and metallicities) for 6 dwarf
elliptical galaxies are presented in this contribution. All of the galaxies 
have the blue absolute magnitude of about $-18$. Their positions on the 
absolute magnitude-surface brightness and absolute magnitude-effective 
radius diagrams classify them as the faint extremity of the ordinary 
elliptical galaxies family, though their effective radii are rather small, 
and at the Kormendy's diagram they are located somewhat between the 
ordinary ellipticals and diffuse elliptical galaxies. All 6 targets 
are members of loose galaxy groups; but 3 of them look rather isolated, 
and 3 of them are projected very close to the bright early-type  
galaxies. We cannot characterize them however as interacting galaxies because
the relative line-of-sight velocities are as high as 300-700 km/s.

\section{Observations and data used}
\label{obs}
I present here some integral-field spectral data
obtained with the Multi-Pupil Fiber Spectrograph (MPFS)
of the Russian 6m telescope and the SAURON data retrieved
from the public ING Archive of the UK Astronomy Data Center.
For the MPFS description, see \cite{mpfsman},
and for the SAURON description, see \cite{sauman}.
Briefly, the former spectrograph gives the field of view of
$16^{\prime \prime} \times 16^{\prime \prime}$  and the spectral 
range of 1500~\AA\ under the spectral resolution
of 3~\AA; the blank sky to subtract from the galaxy spectra
is taken at 4 arcmin from the target. The SAURON possesses 
the field of view of $41^{\prime \prime} \times 33^{\prime \prime}$  
and the spectral range of 550~\AA\ under 
the spectral resolution of 4~\AA; the blank sky is exposed at 1.7 arcmin from 
the center of the field of view. The spaxel is about 1 arcsec in both
spectrograph. NGC~3605 has been observed with the MPFS in
April 2005. All six dwarfs were also exposed with the SAURON in 2007--2008 
in the frame of the project ATLAS-3D. I have calculated the Lick indices
H$\beta$, Mgb, and Fe5270 for every spaxel. The radial profiles of
the Lick indices obtained for NGC 3605 with the MPFS and with the
SAURON agree well. 

\section{Results and Discussion}
\label{finres}
\begin{figure}
\resizebox{1.0\columnwidth}{!}{
\includegraphics{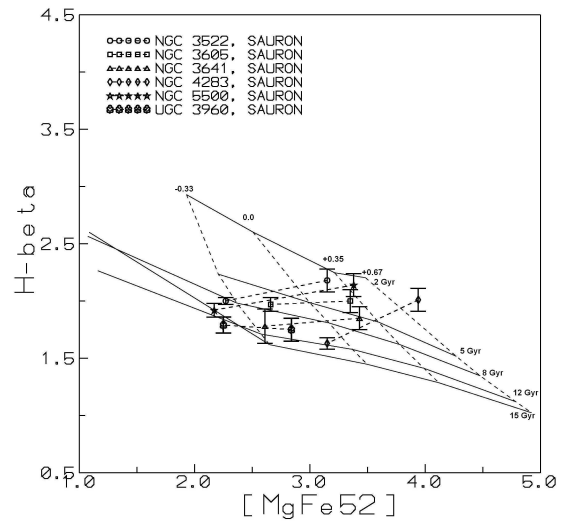} }
\caption{The index-index diagnostic diagram to determine luminosity-weighted
ages and metallicities of the stellar populations. Every pair of signs connected
by dashed straight lines relates to one galaxy; every right reper corresponds to the
nucleus and every left reper is the averaged off-nuclear data taken at 
$R >3^{\prime \prime}$. The labeled solid and dashed lines are constant-age 
and constant-metallicity model sequences by Thomas et al. (2003).}
\label{ii}       
\end{figure}

Figure~\ref{ii} demonstrates the way to determine the luminosity-weighted
(SSP-equivalent) age and metallicities of the stellar populations
in the nuclei and in the main bodies taken outside $R=3^{\prime \prime}$. 
I confront the measured Lick indices, H$\beta$ and the combined magnesium-iron index
[MgFe52], to the population-synthesis models for old SSP by \cite{tmb03}. 
The H$\beta$ of the nucleus in NGC 3522 has been corrected for the weak emission, 
$\Delta \mbox{H}\beta = 0.6 EW(\mbox{[OIII]}\lambda 5007)$. Table~\ref{tabres} 
presents the results of this analysis. In ALL elliptical dwarfs
studied here the nuclei are chemically and evolutionarily decoupled from
the main bodies of the galaxies, the nuclei being younger and more
metal-rich. The metallicity differences between the nuclei and the main 
bodies reach 3-6 times (0.5 - 0.8 dex) while typical enrichments for 
the chemically decoupled nuclei in giant galaxies are known to be around 
2 times (0.3 dex) \cite{me2006}. Evidently, the stellar nuclei have been 
formed dissipatively, in intense secondary star formation bursts. The main 
bodies of all the  dwarfs except NGC~3605 are older than 8 Gyr; in NGC 3605 
the nuclear starburst has been perhaps more extensive than in other galaxies 
that is also implied by asymmetric large-scale distribution of the blue colour
in this galaxy. We may suggest that the young SSP-equivalent
age found by e.g. \cite{caldwell} for nearby dwarf early-type galaxies, 
$\langle T \rangle=3.6$ Gyr, relates precisely to the nuclei, not to the whole 
galaxies. Then the correlation between the mass of a galaxy (stellar velocity 
dispersion) and the age of its stellar population found more than once among 
nearby elliptical galaxies, e.g. by \cite{trager00,howell}, and others, which is
usually treated as a manifestation of `downsizing', is in fact an artifact
caused by the use of the data of aperture spectroscopy centered onto the nuclei.
Minor mergers are frequent events among all types of galaxies, and they
have to provoke nuclear starbursts if the gas is involved; but in a dwarf
galaxy the contribution of the products of nuclear starburst into the
total luminosity of the central part of the galaxy must be more impressive
than in a giant one. So their centers look younger than the centers of giant
ellipticals. If we take the main bodies as representative for the whole
galaxies, the dwarfs would appear as old as the giants.

\section{Acknowledgements}
\label{ack}
This research is based on data obtained from the Isaak Newton Group Archive
which is maintained as part of the CASU Astronomical Data Centre at the Institute
of Astronomy, Cambridge, and also on the data obtained at the Russian 6-m telescope.
The 6m telescope of the Special Astrophysical Observatory is operated under 
the financial support of the Ministry of Science and Education of Russian
Federation (registration number 01-43).

%

%
\begin{table}
\caption{The ages and metallicities for the cores and main bodies}
\label{tabres}       
\begin{tabular}{|l|cc|cc|}
\hline\noalign{\smallskip}
NGC/UGC  &  \multicolumn{2}{c|}{NUCLEUS}     &  \multicolumn{2}{c|}{MAIN BODY 
($R>3^{\prime \prime}$)} \\
       & Age, Gyr  &  [Z/H]   &   Age, Gyr    &    [Z/H] \\
\noalign{\smallskip}\hline\noalign{\smallskip}
N3522  &   $3\pm 1$    &   $+0.2$  &     $8\pm 0.5$    &   $<-0.3$ \\
N3605  &   $4\pm 1$    &   $+0.3$  &     $7\pm 1$      &   $-0.2$ \\
N3641  &   $7\pm 2$    &   $+0.2$  &    $12\pm 3$      &    $-0.3$ \\
N4283  &   $3\pm 0.5$  &   $+0.7$  &    $13\pm 1$      &    $-0.1$ \\
N5500  &   $3\pm 1$    &   $+0.4$  &     $>12$        &     $<-0.3$ \\
U3960  &  $10\pm 2$    &   $-0.2$  &     $>12$        &     $<-0.3$ \\
\noalign{\smallskip}\hline
\end{tabular}
\end{table}

\end{document}